\documentclass[12pt]{article}
\usepackage{epsfig}
\usepackage{subfigure}
\usepackage{lineno}
\def\beq{\begin{equation}}
\def\eeq#1{\label{#1}\end{equation}}
\def\eeqn{\end{equation}}

\def\beqa{\begin{eqnarray}}
\def\eeqa#1{\label{#1}\end{eqnarray}}
\def\eeqan{\end{eqnarray}}

\let\bar=\overbar

\def\Dslash{\not{\hbox{\kern-4pt $D$}}}
\def\dslash{\not{\hbox{\kern-2pt $\del$}}}

\def\msb{{\bar{\ssstyle M \kern -1pt S}}}

\def\Title#1{\begin{center} {\Large {\bf #1} } \end{center}}
\def\fig#1{Fig.~\ref{#1}}
\def\eq#1{Eq.~(\ref{#1})}
\def\tev{\mbox{~TeV}}
\def\gev{\mbox{~GeV}}
\def\gevc{\mbox{~GeV/$c$}}

\def\pt#1{\ensuremath{p_{\rm T#1}}}

\newcommand{\snn} {\ensuremath{\sqrt{s_{NN}}}}

\newcommand{\pp} {\ensuremath{pp}}

\begin{document}

\Title{Particle Correlation Results from the ALICE Experiment at LHC}

\bigskip\bigskip


\begin{raggedright}  

{\it Dong Jo Kim\footnote{Email:djkim@jyu.fi} for the ALICE Collaboration\index{Kim, D.J}\\
University of Jyv\"askyl\"a \& Helsinki Institute of Physics Finland}
\bigskip\bigskip
\end{raggedright}

\begin{abstract}
Measurements of two-particle correlations of inclusive and identified charged particles performed with the ALICE detector in Pb--Pb collisions at \snn = 2.76\tev\ are presented. The near-side jet shape is analyzed in the low \pt{} regions ($1<\pt{}<8\gevc$). While the RMS of the peak in $\Delta\varphi$-direction is independent of centrality within uncertainties, we find significant broadening in $\Delta\eta$-direction from peripheral to central collisions. The near-side $p/\pi$ ratio of particles associated to a trigger particle from jet fragmentation in the central Pb--Pb collisions is consistent with vacuum fragmentation in the measured momentum region ($1.5<\pt{}<4.5\gevc$).

\end{abstract}

\section{Introduction}
In-medium modification of jet fragmentation functions (FF) in heavy ion collisions is thought to be a direct manifestation of the parton energy loss in the medium~\cite{Borghini:2005em,Renk:2008pp,Armesto:2009fj,Zapp:2008gi,Zhang:2009rn}.
Despite of the large parton energy loss signaled by the suppression of high transverse momentum hadrons~\cite{PhysRevLett.106.032301} and the substantial imbalance of the jet transverse energies in di-jet events~\cite{PhysRevC.84.024906},
it has been found that jet FF in Pb--Pb collisions are quite similar to the ones measured in \pp\ collisions~\cite{Chatrchyan:2012gw}.
However there are several caveats in this analysis~\cite{Chatrchyan:2012gw} that should be mentioned, a quite high momentum cut (4\gevc) is imposed on the input particles for the jet reconstruction and a rather small cone size (R=0.3) is used.
For 4\gevc\ hadrons from the model prediction~\cite{Borghini:2005em}, one expects only 20-30$\%$ modification in the yield for a 100\gev\ jet.
Also note that the statistical and systematical errors on the results are of similar magnitude as the expected modification.

Since the radiated energy may result in low \pt{} particles emitted at large angles w.r.t the jet-axis~\cite{PhysRevC.84.024906} and the coupling to the longitudinally flowing medium~\cite{PhysRevLett.93.242301} will lead to broadening that is larger in $\Delta\eta$ than in $\Delta\varphi$,
the motivation for the present analysis in ALICE is to extend the jet shape studies into low \pt{} particles ($\pt{}<8\gevc$) in large $\Delta\eta$ region (wide angular range w.r.t the jet-axis). Furthermore, identified particle ratios associated with the jet and those from the bulk are measured in order to study the jet hadrochemistry~\cite{Sapeta:2007ad} as well as to test coalescence or recombination mechanism~\cite{PhysRevLett.90.202303,Greco:2003xt,PhysRevC.68.034904}.

\section{Detectors and Data Samples}
A description of the ALICE detector and its performance can be found in \cite{Aamodt:2008zz}.
In this analysis, the following detector subsystems were used: the Time Projection Chamber (TPC) \cite{Alme:2010ke}, the Inner Tracking System (ITS), two forward scintillator arrays (VZERO) and the Time-Of-Flight array (TOF)~\cite{Alessandro:2006yt}.
The TPC is the main tracking detector, providing full azimuthal coverage in the pseudo-rapidity range $|\eta| < 1.0$.
The VZERO~detectors cover the pseudo-rapidity ranges $-3.7 < \eta < -1.7$ and $2.8 < \eta < 5.1$.  They are used\ to determine the centrality in  Pb--Pb collisions.
For particle identification (PID) the specific energy loss measured in the TPC as well as the time of flight measured by the TOF system are used. 
About 15 million minimum-bias Pb--Pb events at $\snn$ = 2.76\tev\ and 55 million \pp\ events at $\sqrt{s}$ = 2.76\tev\ are analyzed.

\section{Two Particle Correlations}
The per trigger yield of associated hadrons is measured as a function of the azimuthal angle difference $\Delta\varphi= \varphi_{trig} - \varphi_{assoc}$ and pseudo-rapidity difference
$\Delta\eta=\eta_{trig} - \eta_{assoc}$
\begin{eqnarray}
{1 \over N_{trig}}{d^{2}N_{assoc}\over d\Delta\varphi d\Delta\eta}~|~\pt{,trigg},\pt{,assoc},centrality
\label{eq:corr}
 \end{eqnarray}
 where $N_{assoc}$ is the number of particles associated to the given number of trigger particles $N_{trig}$. This quantity is measured for different ranges of trigger and associated particle transverse momentum in various centrality bins.
The measured distributions are corrected with mixed events to correct for two-track acceptance in bins of centrality and vertex position, where the different $z$ vertex bins are combined by calculating a weighted average as described in the following equations.
First we measure the per trigger yield of associated hadrons in different $z$ vertex bin (~\eq{eq:mixing1}) and weight it by the number of triggers in $z$ bin and calculated the weighted average (~\eq{eq:mixing3}).
The normalization for the mixed events is chosen to be 1 where $\Delta\varphi$ and $\Delta\eta$ is zero  (~\eq{eq:mixing2}).
Also this per trigger yield of associated hadrons are corrected for tracking efficiency and contamination.

\begin{eqnarray}
\label{eq:mixing1}
{d^2N^{raw} \over d\Delta\varphi d\Delta\eta}(\Delta\varphi,\Delta\eta,z) = {1 \over N_{trig}(z)}{N^{same}_{pair}(\Delta\varphi,\Delta\eta,z) \over N^{mixed}_{pair}(\Delta\varphi,\Delta\eta,z)} 
\beta(z)~~~~~~~~~~~~~~~~~~\\
\label{eq:mixing2}
 \beta(z) = N^{mixed}_{pair}(\Delta\varphi=0,\Delta\eta=0,z)~~~~~~~~~~~~~~~~~~~~~~~~~\\
\label{eq:mixing3}
{d^2N \over d\Delta\varphi d\Delta\eta}(\Delta\varphi,\Delta\eta) = {1 \over \Sigma_z N_{trig}(z)} {\Sigma_{z} N_{trigg}(z)} \times {d^2N^{raw} \over d\Delta\varphi d\Delta\eta}(\Delta\varphi,\Delta\eta,z)
 \end{eqnarray}


\section{Results}




\subsection{Near Side Peak Shapes}
\begin{figure}[htbp]
\resizebox{0.30\columnwidth}{!}{\includegraphics{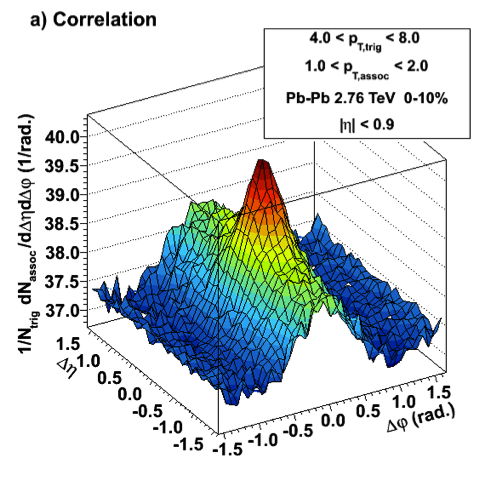}}
\resizebox{0.30\columnwidth}{!}{\includegraphics{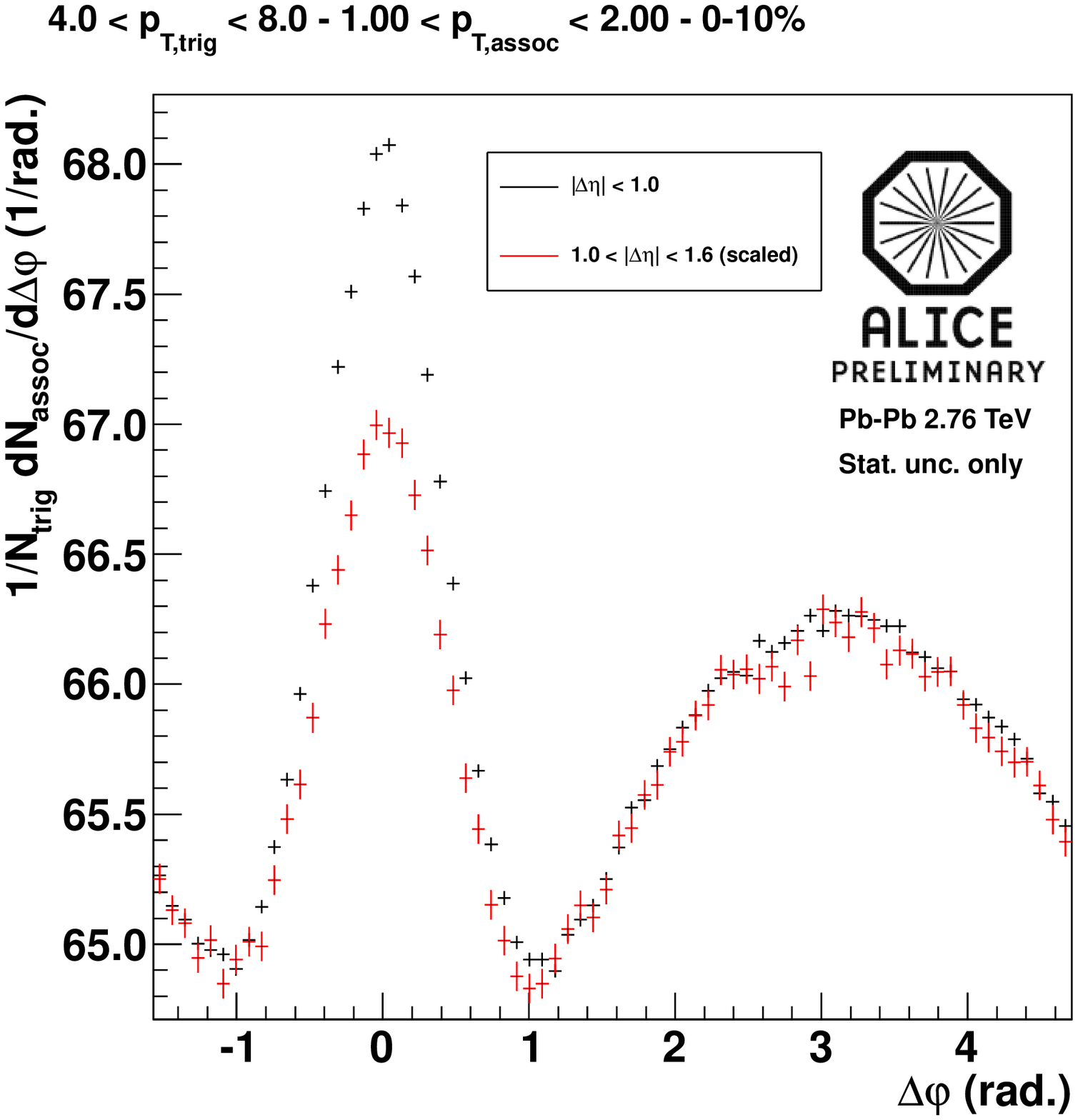}}
\resizebox{0.30\columnwidth}{!}{\includegraphics{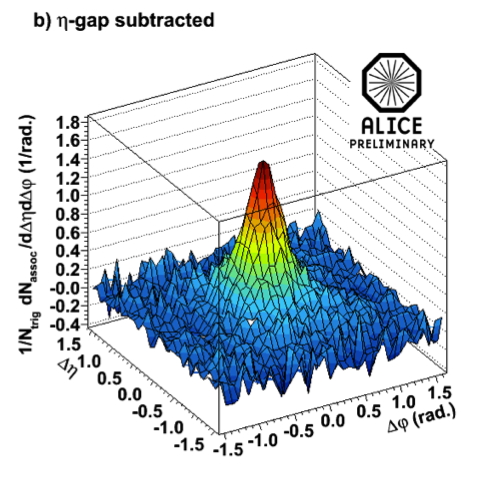}}
\caption{Per trigger yield of associated hadrons before (a) and after (b) $\eta$-gap subtraction. The middle panel shows the projection of the per trigger yield of associated hadrons (a) to $\Delta\varphi$ in $|\Delta\eta|<1$ (black) and $|\Delta\eta|>1$ (red).}
\label{fig:cfphibck}
\end{figure}

\begin{figure}[htbp]
\resizebox{1.0\columnwidth}{!}{\includegraphics{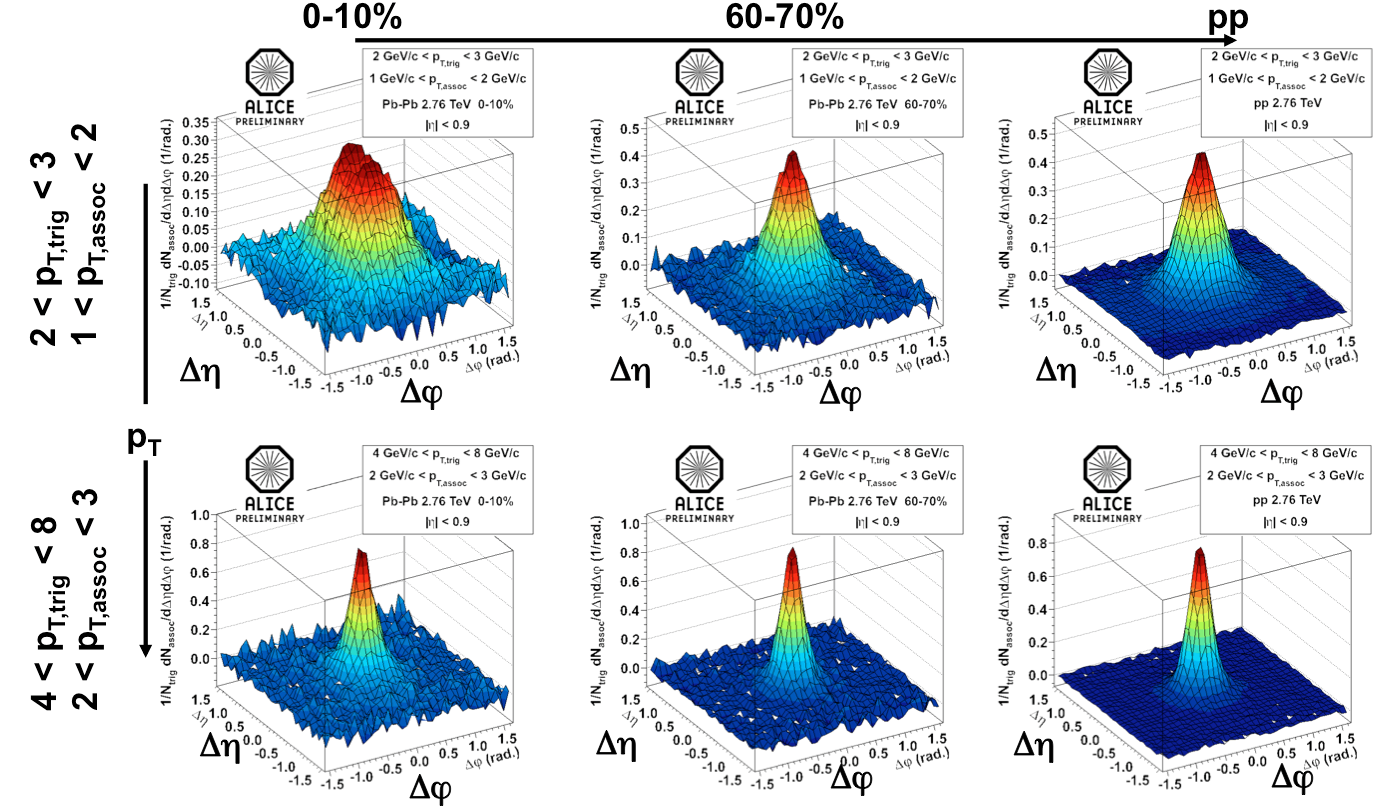}}
\caption{Evolution of the per trigger yield of associated hadrons for two different trigger and associate particle momentum ranges in central, peripheral and \pp\ collisions. Trigger and associated particle momentum range and centrality selection is shown on the figure. \pp\ data is on the right panel. 
}
\label{fig:cfphi}
\end{figure}

The near side jet shapes are studied in low momentum ranges of trigger and associated particles.
\fig{fig:cfphibck}(a) shows a two dimensional "per trigger yield of associated hadrons" in $\Delta\varphi$ and $\Delta\eta$ space in central (0-10\%) Pb--Pb collisions. The trigger particle momentum is from 4 to 8\gevc\ and associated particle transverse momentum from 1 to 2\gevc. 
We have estimated the background yields from $\Delta\eta$-independent sources such as flow in the region of $|\Delta\eta|>1$.
The middle panel in \fig{fig:cfphibck} shows the 1-dimensional projection of \fig{fig:cfphibck}(a), where the flow background, long range correlation $|\Delta\eta|>1$ (normalized by the acceptance), is shown in red and the mix of short and long range correlation in black ($|\Delta\eta|<1$).
In the near side, we see a clear excess of correlated jet signal above the background.
Once we subtract the long range correlation from the region $|\Delta\eta|<1$,  we obtain the final \fig{fig:cfphibck}(b)  where we see a clear jet peak.
We repeat this procedure in different trigger, associated particle momentum bins as well as various centrality bins. The results are shown in \fig{fig:cfphi}. Peripheral and \pp\ look similar and we observe wider distribution in central collisions as compared with the peripheral and \pp\ collisions.

\subsection{Jet Shape Characterization}
\begin{figure}
\centering
\subfigure[$\sigma_{\Delta\varphi}$]{\includegraphics[width=0.45\linewidth]{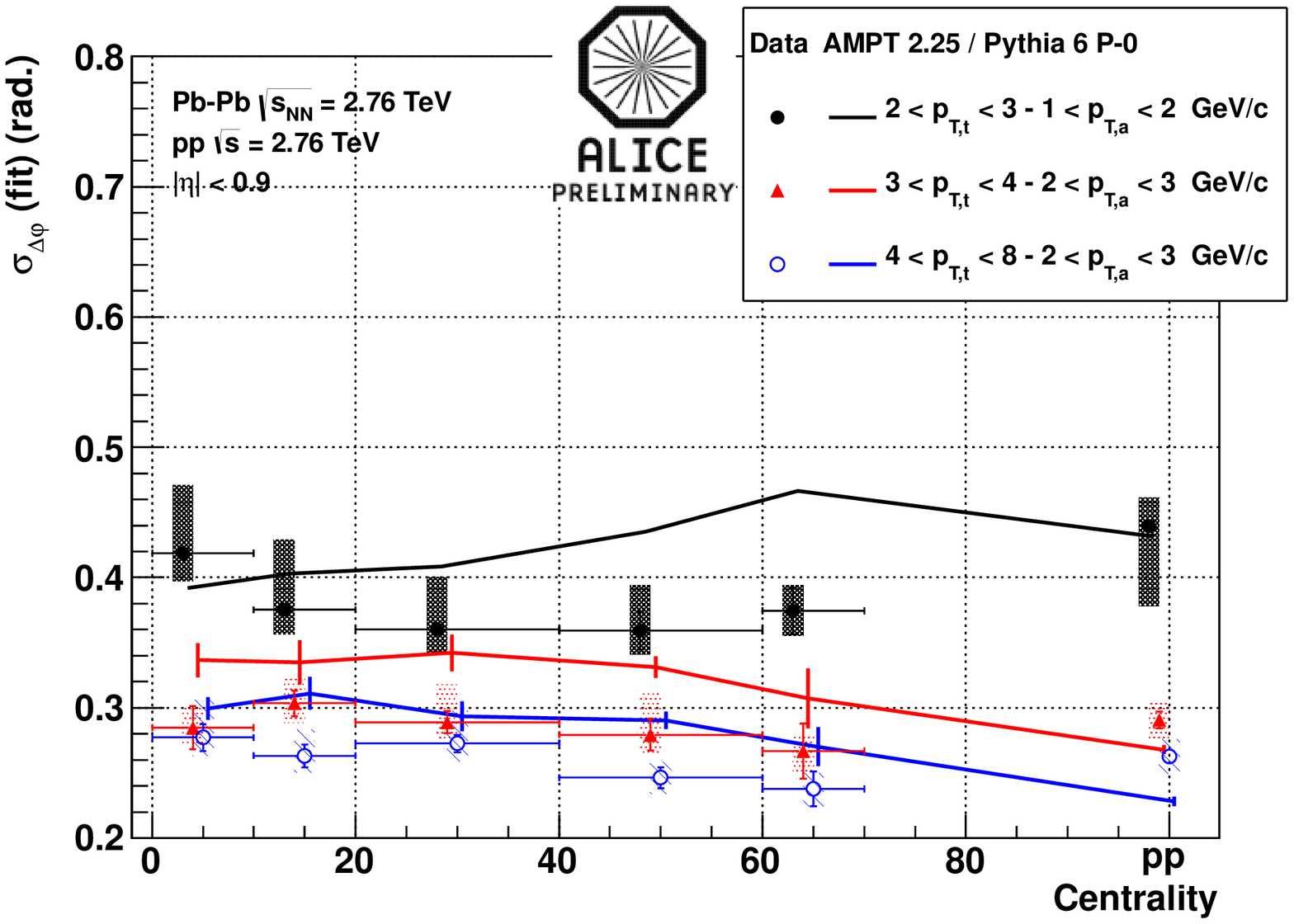}}
\subfigure[$\sigma_{\Delta\eta}$]{\includegraphics[width=0.45\linewidth]{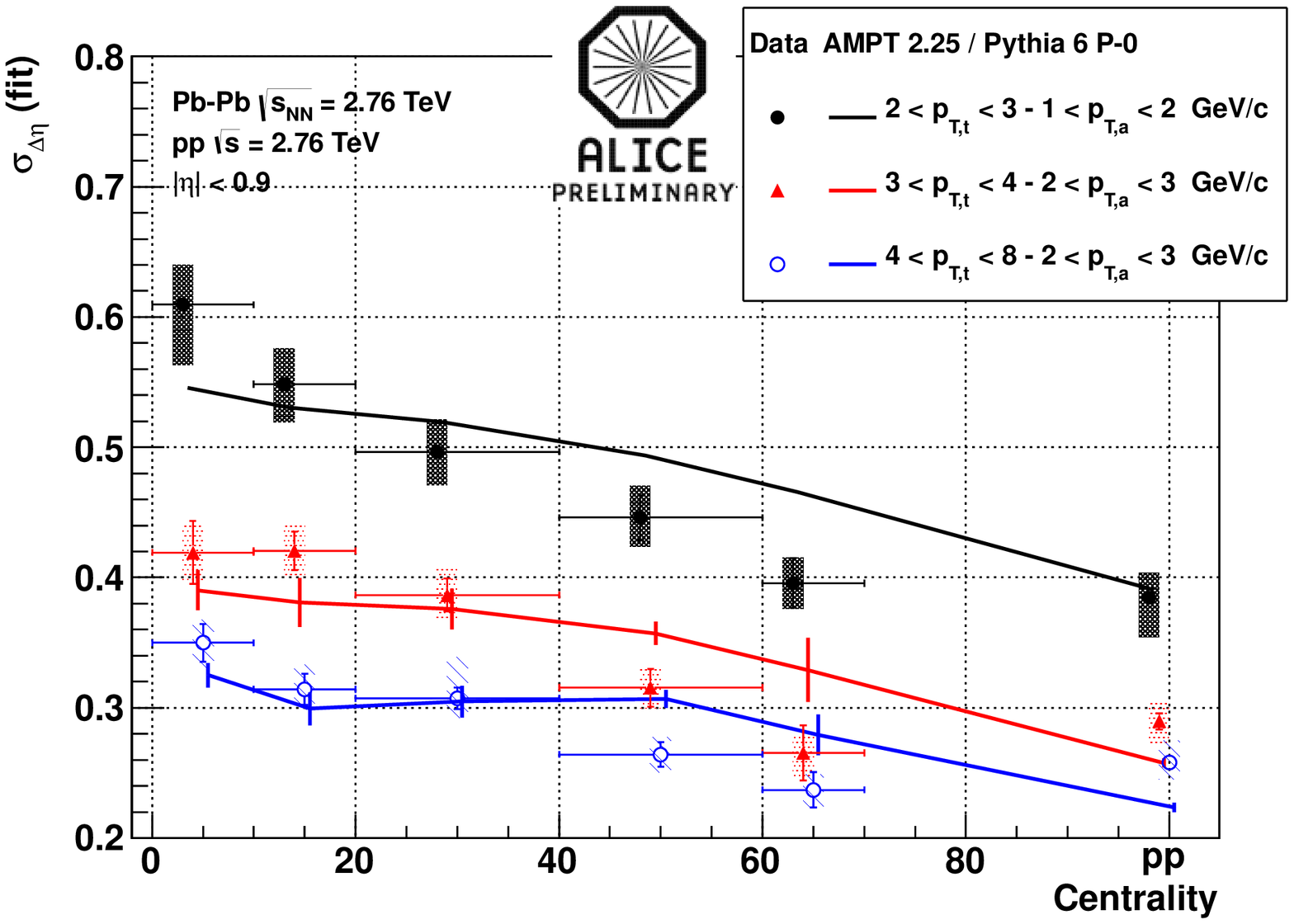}}
\caption{Centrality dependence of $\sigma_{\Delta\varphi}$ (left) and  $\sigma_{\Delta\eta}$  (right) for three different bins of \pt{,trigg} and \pt{,assoc}. The lines are extracted from AMPT~\cite{Lin:2004en,Xu:2011fi} (A MultiPhase Transport Code; version 2.25 with string melting in PbÐPb) and Pythia~\cite{Sjostrand:1993yb} simulations (pp).}
\label{fig:peakrms}
\end{figure}

The jet shape is quantified with RMS and excess kurtosis which is a measure of its peakedness.
The near side peak was fitted with a sum of  two 2 dimensional gaussians.
The obtained results, $\sigma_{\Delta\eta}$ and $\sigma_{\Delta\varphi}$ are shown in \fig{fig:peakrms} as a function of centrality for various trigger and associated particle momentum bins.
The last data point in the centrality axis corresponds to the result extracted from \pp\ collisions.

\begin{figure}
\centering
\subfigure[${\Delta\varphi}$]{\includegraphics[width=0.45\linewidth]{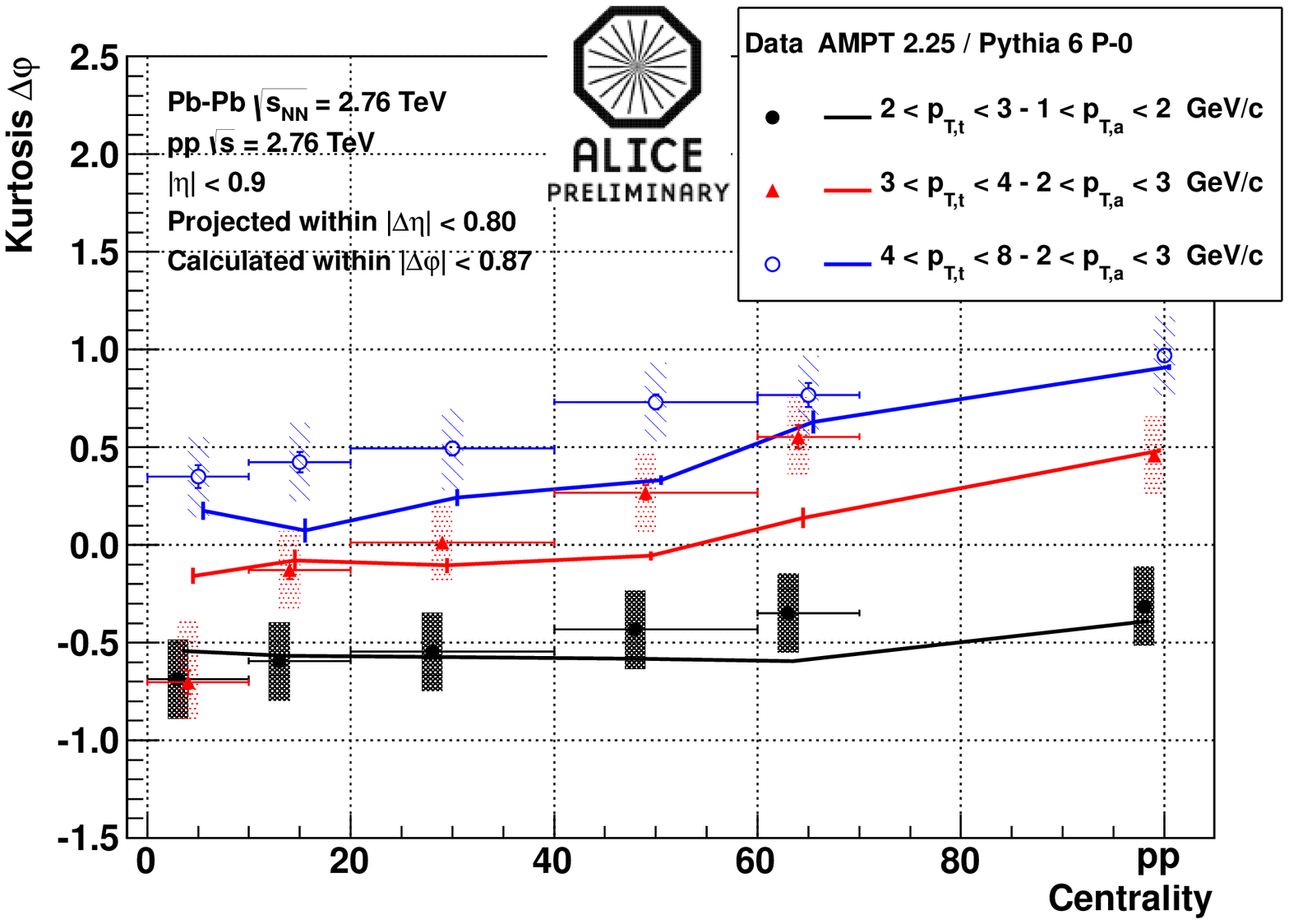}}
\subfigure[${\Delta\eta}$]{\includegraphics[width=0.45\linewidth]{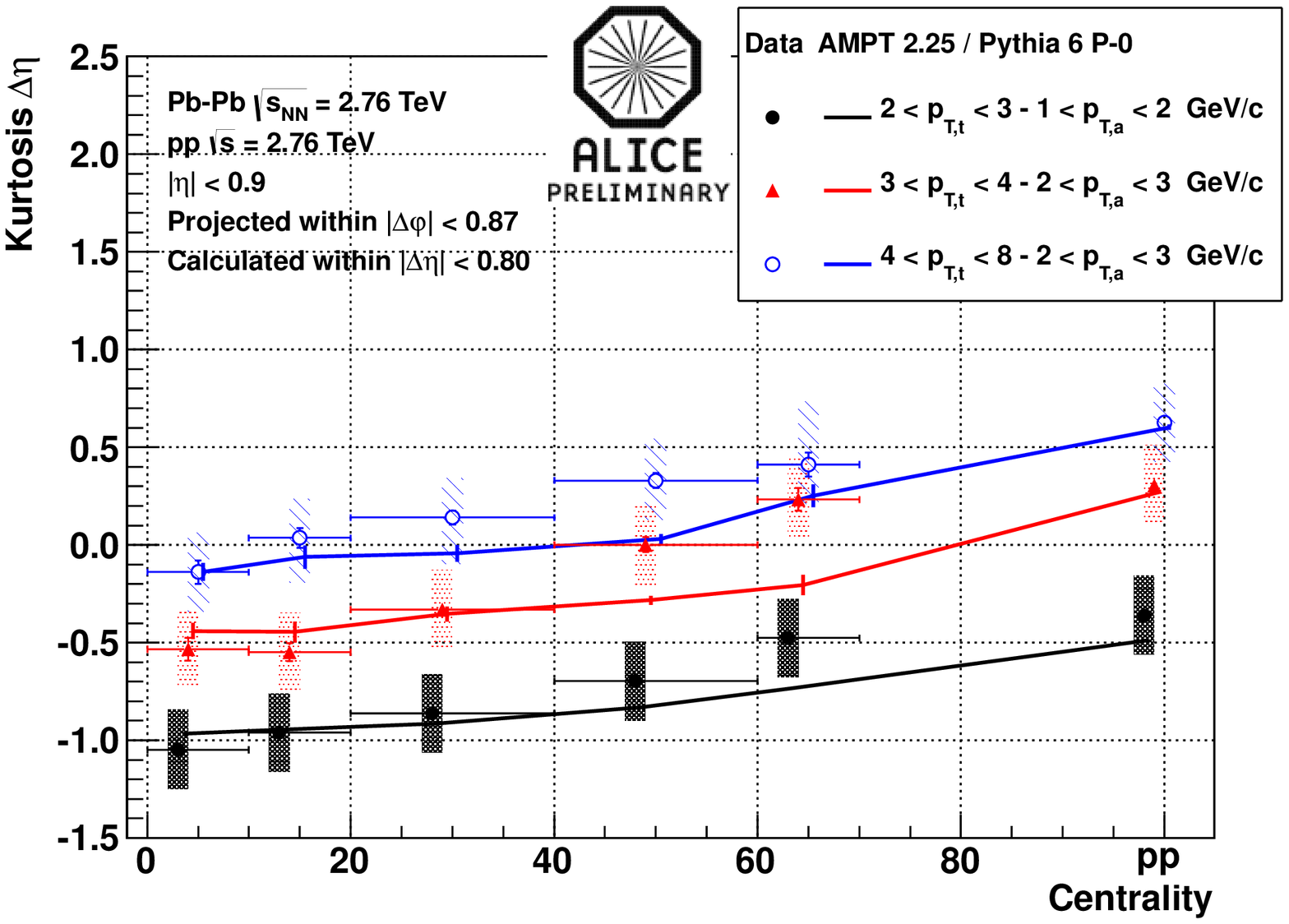}}	
\caption{Centrality dependence of the kurtosis in ${\Delta\varphi}$ (left) and  ${\Delta\eta}$  (right) for three different bins in \pt{,trigg} and \pt{,assoc}. The lines represent AMPT calculation~\cite{Lin:2004en,Xu:2011fi} for the Pb--Pb collisions and Pythia~\cite{Sjostrand:1993yb} simulations for \pp.}
\label{fig:dphietakurtosis}
\end{figure}

As expected,  $\sigma_{\Delta\eta}$ and $\sigma_{\Delta\varphi}$ in higher \pt{} bins are smaller than low \pt{} bins since the jet is more collimated.
But while we do not see any significant centrality dependence of $\sigma_{\Delta\varphi}$, we observe a significant increase of  $\sigma_{\Delta\eta}$ towards central events
for every particle momentum bin shown in this figure.
For the lowest \pt{} bins,  the eccentricity ($(\sigma_{\Delta\eta}-\sigma_{\Delta\varphi})/(\sigma_{\Delta\eta}+\sigma_{\Delta\varphi}))$ is about 0.2.
The evolution from peripheral to \pp\ is smooth. \fig{fig:dphietakurtosis} shows the excess kurtosis in same trigger and associated particle momentum bins as a function of centrality.
A clear \pt{} dependence of the excess kurtosis is seen and it increases with \pt{}. The excess kurtosis decreases from \pp\ to peripheral and central events. The RMS and excess kurtosis of the near-side peak are well reproduced by the AMPT model~\cite{Lin:2004en,Xu:2011fi} shown in lines both in \fig{fig:peakrms} and  \fig{fig:dphietakurtosis}.

\begin{figure}
\centering
\subfigure[$\eta-gap$ subtracted]{\includegraphics[width=0.45\linewidth]{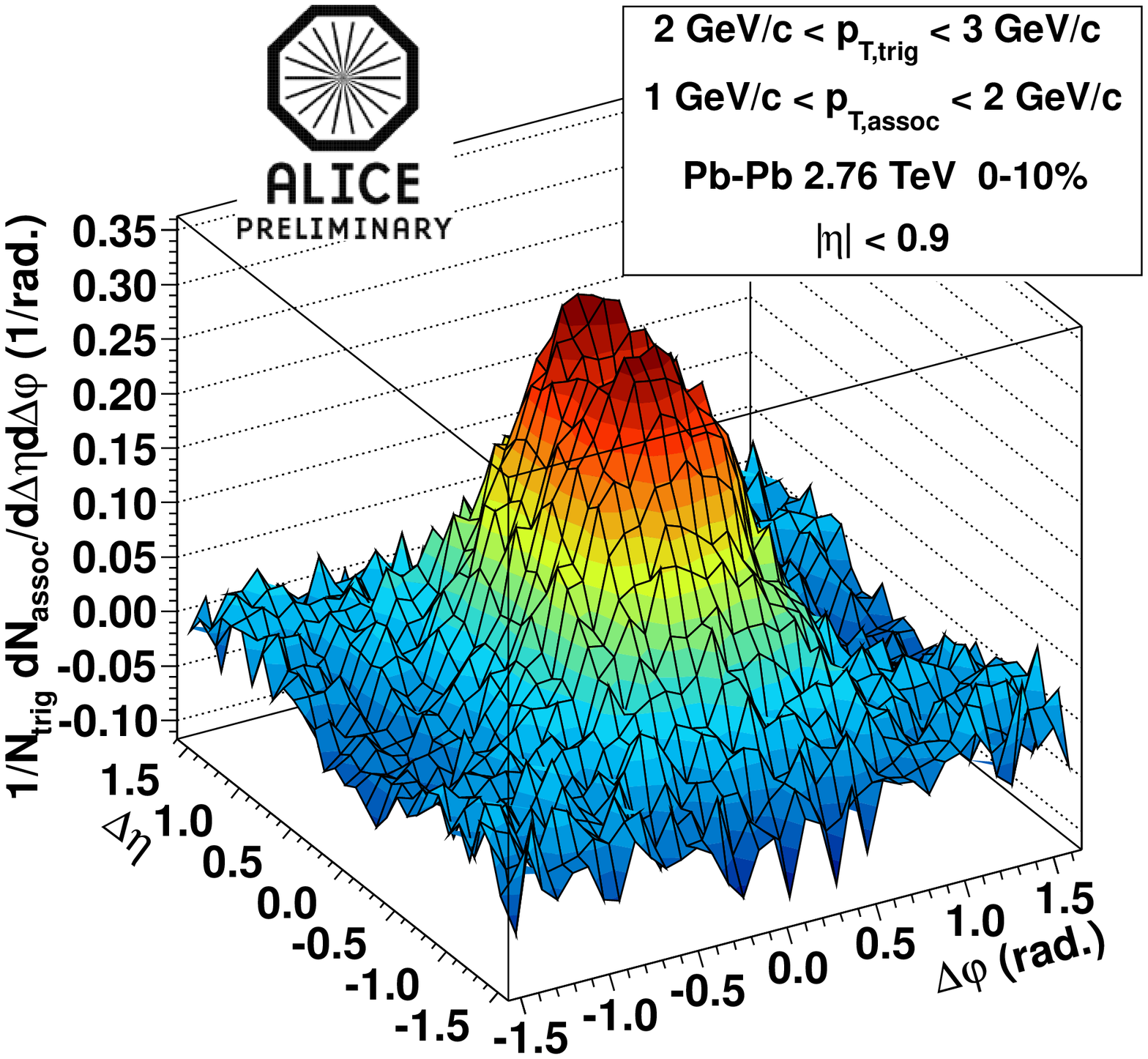}}
\subfigure[$\Delta\eta,\Delta\varphi$ projection]{\includegraphics[width=0.45\linewidth]{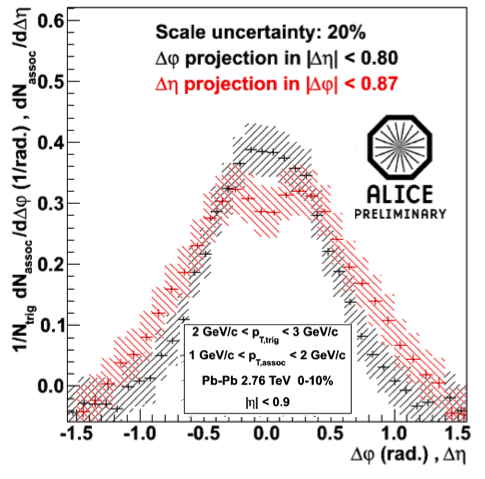}}	
\caption{Per trigger yield of associated hadrons in $\Delta\varphi$ and $\Delta\eta$ in 0-10\% central collisions for $2<\pt{,trigg}< 3$ and $1<\pt{,assoc}<2$ \gevc.}
\label{fig:lowptrms}
\end{figure}

In the lowest \pt{} bin shown in \fig{fig:lowptrms} ( $2<\pt{,trigg}< 3$, $1<\pt{,assoc}<2$ \gevc) in 0-10\% central collisions, one sees a structure with a flat top in $\Delta\eta$ while this is not seen in $\Delta\varphi$.
This is a surprising result which might be explained by the interplay of jets with the flowing bulk~\cite{PhysRevLett.93.242301,Lin:2004en,Xu:2011fi}. 

\subsection{$p/\pi$ Ratio in Bulk and Jet Region}
Baryon over meson ratios differ significantly between heavy ion and \pp\ collisions which might be attributed to radial 
flow and coalescence or recombination mechanism~\cite{PhysRevLett.90.202303,Greco:2003xt,PhysRevC.68.034904}.
Few years back, S. Sapeta, U.A Wiedemann have studied the jet fragmentation in heavy ion collisions for various particle species~\cite{Sapeta:2007ad}. They show that medium-modification of the parton shower can result in significant changes in jet hadrochemistry~\cite{Sapeta:2007ad}, e.g, strong enhancement of proton compared to pion yield in the medium is expected from this model.
ALICE has utilized the correlation method to measure the $p/\pi$ ratio in Pb--Pb collisions where this ratio is measured both in the jet region and bulk region separately.

\begin{figure}[htbp]
\resizebox{0.50\columnwidth}{!}{\includegraphics{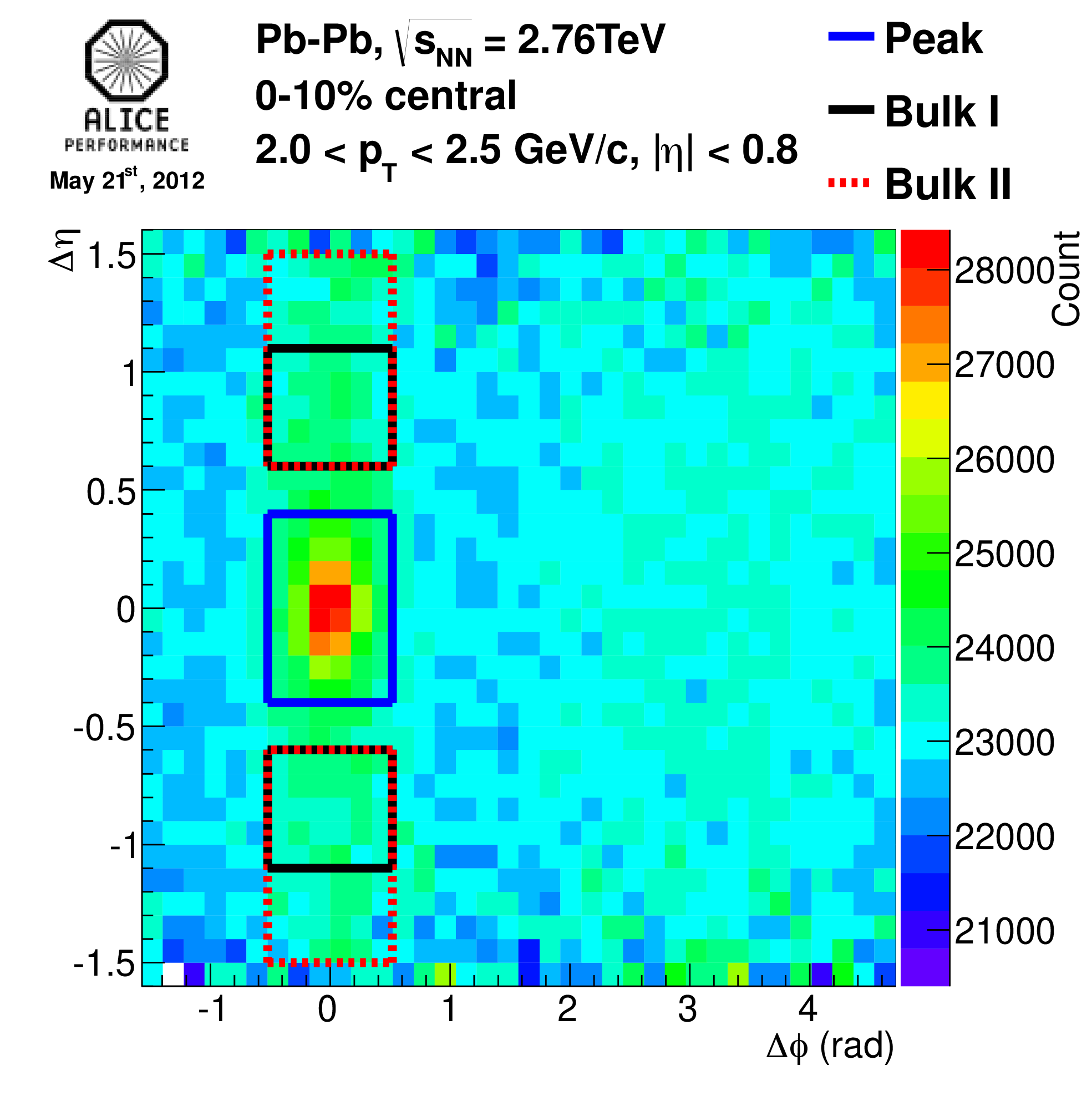}}
\resizebox{0.50\columnwidth}{!}{\includegraphics{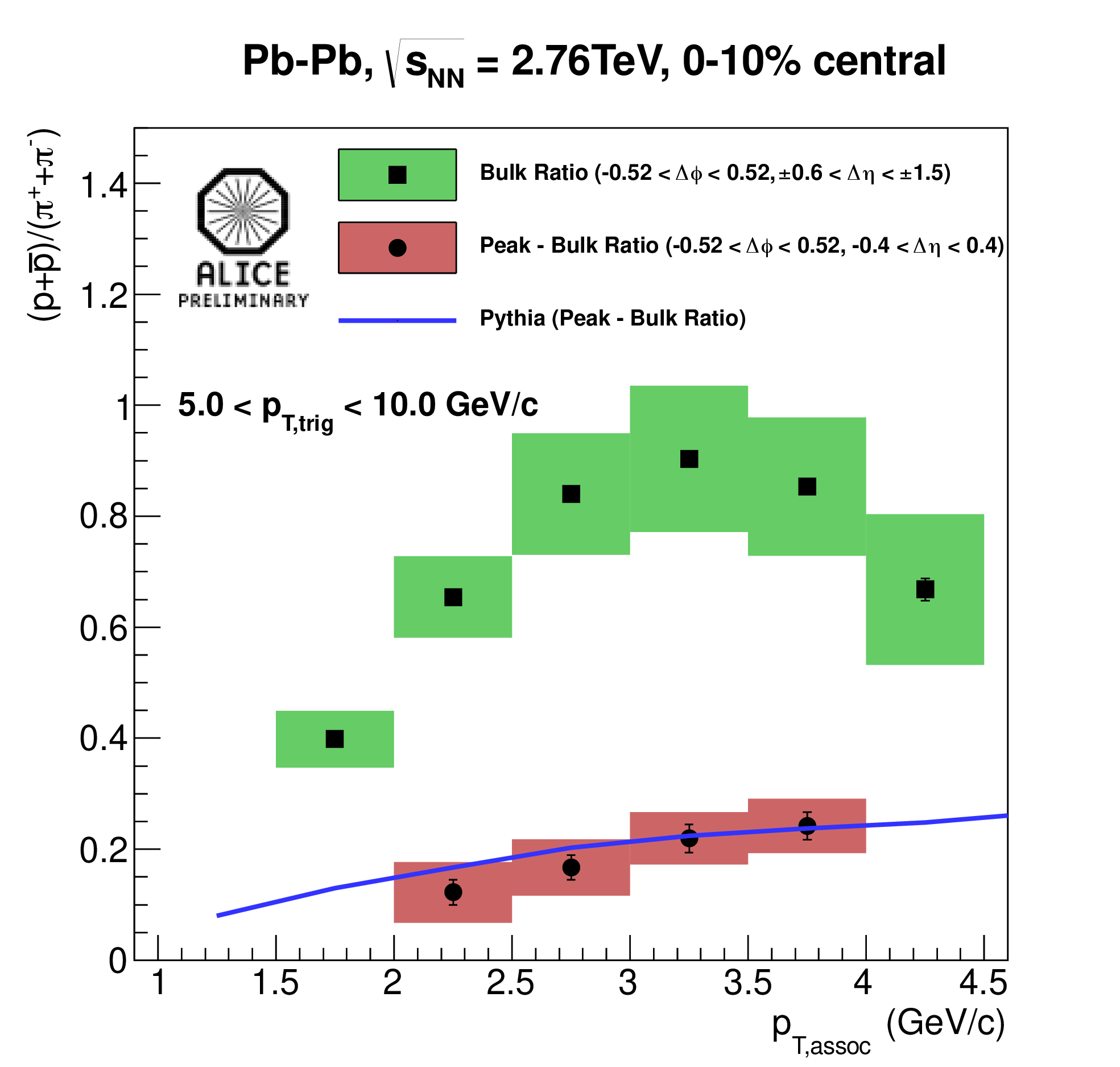}}
\caption{Left panel: Jet and bulk regions. Right panel: $p/\pi$ ratio in the bulk (squares) and peak-bulk (circles) regions compared to the PYTHIA (line)}
\label{fig:pidjetbulk}
\end{figure}
	
In this analysis, 5 to 10\gevc\ charged hadrons are used as the trigger particles. The associated particles are identified as $\pi$ or proton in the momentum range from 1.5 to 4.5\gevc.
Combined particle identification with specific energy loss in the TPC and time of flight in the TOF was used.
\fig{fig:pidjetbulk} (left panel) shows the two dimensional $\Delta\varphi$ and $\Delta\eta$ per trigger yield of associated hadrons for associated particle momentum range from 2.0 to 2.5 \gevc.
Jet (peak) region is chosen in the range, $-0.4<\Delta\eta<0.4,-0.52<\Delta\varphi<0.52$ in radian and bulk region is selected in long range correlation dominated region as discussed in the previous section.
For those associated particles in the selected regions, the PID method is applied to get the yields of each species. These yields are corrected for tracking and PID efficiency. 
No correction for feeddown from e.g. $\Lambda$ decay has been applied.
Once the yields are obtained both in the jet region and the bulk region, the yields for each species in the bulk region is subtracted from yields of the jet region to get final yields originating only from jets.
The normalization factor for the jet region from the bulk region is given by the acceptance difference between the jet and bulk region.

The resulting $p/\pi$ ratio in the bulk region in 0-10\% central collisions is shown in \fig{fig:pidjetbulk} on the right panel. It exhibits strong enhancement of protons for higher \pt{} and it is consistent with the inclusive measurement~\cite{Floris:2011ru}.
However, the ratio in the jet region is much smaller than in the bulk region.
While we do not have the same result in \pp\ collisions yet, the ratio in jet region is compared with PYTHIA (pythia6.4) ~\cite{Sjostrand:1993yb} for \pp\ collisions.
Both results agree within the errors. 
This indicates no significant change of near side jet hadrochemistry in the central Pb--Pb collisions in the measured momentum region (1.5 to 4.5\gevc).

\section{Summary}
Two-particle correlations have been used to quantify the signatures of the modified jet fragmention in the medium in the low \pt{} regions ($1<\pt{}<8\gevc$) in Pb--Pb collisions at \snn = 2.76\tev. While the RMS of the near-side peak in $\Delta\varphi$-direction is independent of centrality within uncertainties, we find significant broadening in $\Delta\eta$-direction from peripheral to central collisions. 
The RMS and excess kurtosis of the near-side peak are well reproduced by the AMPT model, which might explain the interplay of jets with the flowing bulk~\cite{Lin:2004en,Xu:2011fi}.
The near-side $p/\pi$ ratio of particles from jets in central Pb--Pb collisions is similar as that of PYTHIA~\cite{Sjostrand:1993yb}. This indicates no significant change of near side jet hadrochemistry in the central 0-10\% Pb--Pb collisions compared with \pp\ collisions in the measured momentum region ($1.5<\pt{}<4.5\gevc$).


\def\Discussion{
\setlength{\parskip}{0.3cm}\setlength{\parindent}{0.0cm}
     \bigskip\bigskip      {\Large {\bf Discussion}} \bigskip}
\def\speaker#1{{\bf #1:}\ }
\def\endDiscussion{}


 
\end{document}